\documentstyle{l-aa} 
\input epsf.sty

\begin{document}
   \thesaurus{13(13.18.1); 
              11(11.09.1;       
           11.19.6;      
               11.17.4; 
                 11.14.1)} 

   \title{Parsec-scale structures of radio galaxies in the 2-Jy sample}

\author{T. Venturi
\inst{1}, R. Morganti\inst{2}, T. Tzioumis\inst{3},
J. Reynolds\inst{3}}

\offprints{T. Venturi(tventuri@ira.bo.cnr.it)}

\institute{1. Istituto di Radioastronomia del CNR, Via Gobetti 101, 
40129 Bologna, Italy\\
2. Netherlands Foundation for Radio Astronomy, Postbus 2, AA 7990 Dwingeloo,
The Netherlands\\
3. ATNF-CSIRO, P.O. Box 76, Epping, NSW 2121, Australia}

\date{Received ...; accepted ...}
   \maketitle

\markboth{T.Venturi et al.: Parsec-scale structures of 2-Jy radio galaxies}
{T.Venturi et al.: Parsec-scale structures of 2-Jy radio galaxies}

   \begin{abstract}
In this paper we present the results of VLBI observations
of six radio galaxies belonging to the 2-Jy sample.
The selected objects are 3C17, PKS 0620$-$52, PKS 0625$-$35,
PKS 1318$-$43, PKS 1333$-$33 and 3C317.
The first is a high power radio galaxy and the remainder are all 
low power objects.
Our observations were carried out with a set of different arrays
and frequencies, and cover a range of resolutions from
a few mas to a few tens of mas.

Parsec-scale images are presented and discussed in the
light of unification  for radio loud galaxies.
Estimates for the intrinsic plasma speeds in these objects 
in the proximity of the parsec-scale core and their
orientation to the line of sight are in agreement with
the predictions from unified models and with their
optical and X-ray properties.

We will use H$_o$ = 50 km s$^{-1}$Mpc$^{-1}$,  
q$_o$ = 0 and $S \propto \nu ^{-\alpha}$ throughout the paper.

   \end{abstract}

%

\section{Introduction}

The study of the nuclear regions in radio galaxies with Very Long Baseline
Interferometry (VLBI) has proved to be a very important tool for the
understanding of the properties of the central engine in low and high power
radio galaxies.  Furthermore it has provided a major contribution in 
testing the unified schemes which attribute the main differences 
between the various classes of AGNs to relativistic beaming and 
orientation effects. 

VLBI observations of complete samples of radio galaxies, together with
detailed observations on individual objects (see for example Feretti et al. 
1993; Venturi et al.  1993 and 1995; Giovannini et al.  1994, 1998, 1999; Lara
et al.  1997 and 1999; Jones \& Wehrle 1997; Krichbaum et al.  1998; Taylor
1996), show that lobe-dominated, edge-darkened low power 
FRI radio galaxies and the lobe-dominated,
edge-brightened high power FRIIs (Fanaroff \& Riley 1974)
are almost indistinguishable on the parsec scale. 

Both classes of radio galaxies are typically characterised by one-sided
parsec-scale morphology.  A few cases of symmetric two-sided morphologies are
found both among FRIs, i.e.  Hydra A (Taylor 1996) and 3C338 (Giovannini et
al.  1998), and among FRIIs, for example 3C452 (Venturi et al.  2000,
Giovannini et al., in preparation). 
From these studies, 
relativistic bulk motion of the radio emitting plasma in the radio jets, with
Lorentz factors ranging from a few units up to 10 - 15 were derived. 
 
For both classes of objects it is estimated that the radio emission is
oriented at intermediate to large angles to the line of sight.  This is in
agreement with predictions from the standard unified models where FRII
radio galaxies
represent the misaligned population of flat spectrum quasars (Barthel
1989), while FRIs are the misaligned population of BL-Lac
objects (Urry \& Padovani 1995).  For the latter unification, 
there remain some issues to be clarified. 
For example, on the basis of statistical arguments and optical data,
Chiaberge et al.  (1999, 2000) and Capetti \& Celotti (1999) 
concluded that the Lorentz factors needed for FRIs and BL-Lacs, in order to
account for the different orientation dependent properties, are 
in the range $\gamma \sim~3 - 7$. These values are lower  than derived by 
independent arguments, such as the modelling of the spectral energy
distribution in BL-Lacs (Ghisellini et al. 1998).

Chiaberge et al.  (2000) proposed that a two-phase jet in FRIs could
account for the apparent discrepancy in the standard unification model.
Furthermore they 
underlined the importance of multifrequency studies of the nuclear regions of
radio galaxies and their parent aligned population, in order to overcome the
remaining discrepancies between theoretical predictions and observations, and
to probe the inner parsec-scale regions in AGN. 

In addition, radiative inefficient accretion models, 
such as the Advection
Dominated Accretion Flow (ADAF), have been recently  suggested to 
explain the spectral properties
of some radio galaxies.  In particular, they have been applied to galaxies
with high hard X-ray excess and moderate radio luminosity
(see e.g.  Di Matteo et al. 1999) but no sign of AGN activity from the 
optical lines.  Milliarcsec resolution radio observations 
provide parsec-scale imaging of the innermost regions in radio galaxies,
and allow derivation of the radio spectrum of the nuclear component, 
both crucial pieces of information to test and constrain
the ADAF accretion process.

\medskip 
Given the importance of multi-wavelength information,
as emphasized above, we have started a program to provide
VLBI information for some of the radio sources in the 2-Jy sample.  For the
sources in this sample a wealth of radio, optical, X-ray 
data are already available. 
In this paper we present VLBI radio observations 
of the nuclear region in six
radio galaxies selected from this sample not yet imaged at parsec-scale
resolution, and discuss them in the light
of their large scale radio properties, optical and X--ray features.  In
Section 2 we illustrate the selection criteria and give an overview of the
selected sources; the observations and data reduction are described in
Section 3; individual sources
are discussed in Section 4.  Discussion and conclusions are given in
Section 5.

\section{ Selected radio galaxies from the 2-Jy sample}

Radio sources belonging to the 2-Jy catalogue of radio sources (Wall \&
Peacock 1985) are potentially a very useful tool for the understanding of the
nuclear properties of radio galaxies.  The 2-Jy sample is well studied in a
wide range of wavelengths, from radio (Morganti et al.  1993, 1999 and 
references therein), to optical spectroscopy (Tadhunter et al.  1993, 1998 
and references therein), up to X--ray energies (Siebert et al.  1996, 
Trussoni et al.  1999, Padovani et al. 1999). 
Recently, 23 galaxies in the sample have also been searched
for HI absorption (Morganti et al. 2000, submitted). 
This allows comparative studies, which are crucial to
understand the physics of the inner parsec-scale region of AGN, and
to test unified models and evolutionary scenarios. 

The 2-Jy sample includes radio sources with flux density 
S$_{2.7 GHz} \ge$ 2 Jy, redshift z $<$ 0.7 and declination 
$\delta \le +10^{\circ}$.  Given the frequency of observation,
the sample is biassed towards beamed radio sources.

\noindent
Among the 2-Jy sample we have selected all sources
with S$_{core}$ (5 GHz) $\ge$ 200 mJy not yet imaged at VLBI resolution.  The
core flux density limit was chosen so as to be able to image the
southernmost radio galaxies with the sensitivity of the southern Long Baseline
Array (LBA) at the time the original proposals were made.  In this paper we
present the results of parsec-scale observations for six radio galaxies from
the selected sample.  The list of objects is given in Table 1, where the most
relevant parameters, i.e.  redshift z, core power logP$_c$ 
at 5 GHz (data taken from Morganti et al. 1993)
and total power logP$_t$ at 408 MHz are also reported.  

Among them we have four classical FRIs, one core-halo source and one 
intermediate type FRI/FRII. 
The power range of 3C17 is typical of FRIIs, however its 
arcsecond scale morphology is more similar to the transition
objects between FRIs and FRIIs than to the typical classical
doubles (Morganti et al. 1993; Morganti et al. 1999). 
It is the most distant object among the sources in the present sample.
The total power for the five remaining
galaxies is that of the most powerful FRIs, and their large
scale radio emission is typical of FRI 
except for 3C317, which is classified as core-halo.

%
\begin{table*}

\caption{The sources in the VLBI sample}

 \begin{tabular}{llcrccccl} \hline
Source & Other Name & RA. (1950) &  Dec. (1950) &  z  &  
S$_c$ (5 GHz) & logP$_c$ (5 GHz) & logP$_t$ (408 MHz) & Type\\ 
  &   &  &  &  & Jy &  W Hz$^{-1}$ & WHz$^{-1}$  & \\ \hline   
0035$-$02 & 3C 17    & 00 35 47.17 & $-$02 24 09.3 & 0.220 & 0.662 & 
26.21 & 27.55$^1$ & FRI/II \\ 
PKS 0620$-$52 &      & 06 20 36.85 & $-$52 40 00.9 & 0.051 & 0.260 & 
24.49 & 25.90$^1$ & FRI \\
PKS 0625$-$35 & OH 342 & 06 25 20.22 & $-$35 27 22.0 & 0.055 & 0.600 & 
24.91 & 26.02$^1$ & FRI \\
PKS 1318$-$43 &NGC 5090& 13 18 17.40 & $-$43 26 33.4 & 0.011 & 0.580 & 
23.49 & 24.32$^{\star}$ & FRI \\
PKS 1333$-$33 &IC 4296 & 13 33 47.16 & $-$33 42 39.8 & 0.013 & 0.297 & 
23.35 & 25.41$^2$ & FRI \\
1514$+$07     & 3C 317 & 15 14 16.99 & $+$07 12 17.1 & 0.035 & 0.391 & 
24.34 & 26.12$^1$ & C/H\\ \hline
\end{tabular}

\medskip
Notes to Table 1.

References: $^1$ Large et al., 1981; $^2$ Wright \& Otrupcek 1990.

$^{\star}$ No 408 MHz power is available in the literature. See Sect. 5 in 
text.

\end{table*}

PKS 0625$-$35, PKS 1333$-$33 and 3C317 are all located in Abell
clusters, respectively A3392, A3565 and A2052; PKS 1318$-$43
is located in a low density environment and it is 
interacting with the peculiar companion galaxy NGC5091.
Very little information is available on the galaxy cluster
surrounding PKS 0620$-$52; finally the large scale environment of 3C17
is unknown.

\section{Observations and data reduction}

Our VLBI observations were carried out at various frequencies,
with different array configurations and recording modes. 
The logs of the observations,
together with the most relevant information, are given in Table 2.
The bandwidth of the MK2 observations is 2 MHz; 
the S2 observations were carried out with a single 16 MHz band; 
the MK3 observations were carried out in mode E,
with 7 IFs and a total bandwidth of 14 MHz. 

The initial part of the data reduction (a-priori
amplitude calibration and phase correction) was carried out
differently for the various experiments, given the different
formats of the correlator output; self-calibration and
imaging were performed by means of the NRAO AIPS package 
for all sources. Data correlated 
with the 5 station Block 0 MK2 correlator in Bologna were
written into merge format and read into the Caltech package
(Pearson 1991) for the a-priori calibration;
both the S2 data correlated in Epping and 
the MK3 data correlated in Bonn were read into AIPS, 
where fringe-fitting, self-calibration and imaging were carried out. 

Because of the different array configurations and sensitivities,
the resolution and quality of our final images vary from 
case to case. The flux density scale is accurate within 
a few percent for 3C17 (MK2 - VLBA observations)
and for 3C317 (EVN+Merlin MK3 observations), while for all
sources observed with LBA the uncertainties are of the order 
of 10 - 15\%.

%
\begin{table*}
\caption{Logs of the VLBI observations}

\begin{tabular}{lcccccrc} \hline
Source &   Date     &  Array  &  $\nu$ & Min.- Max. Baseline   & Mode &
Duration & Correlator \\ 
       &            &         &   GHz  &  M$\lambda$         &      &
  hr   &     \\ \hline  
3C17   &   02/10/94 & VLBA$^1$ & 4.99  & 7.3 - 145  & MK2  &
   8   &  Bologna \\
PKS 0620$-$52& 22/06/96 &  LBA$^2$ & 2.29  & 0.7 - 10.3 & S2   & 
  8   &  Epping  \\
PKS 0625$-$35& 23/02/94 & LBA$^3$ & 2.30 & 0.7 - 10.7 & MK2 & 
  8   &  Bologna \\
PKS 1318$-$43& 24/02/94 & LBA$^3$ & 2.30 & 0.4 - 10.6  & MK2 & 
  8   &  Bologna \\
PKS 1333$-$33& 23/06/96 & LBA$^2$ & 2.29 & 0.8 - 10.7 & S2  & 
  8   &  Epping  \\
3C317    & 14/02/95 &EVN+MERLIN$^4$&1.66 & 0.1 - 36.5  &MK3-E& 
 10   &  Bonn    \\
3C317    & 13/05/96 &EVN+MERLIN$^5$&4.99 & 0.03 - 12.5  &MK3-E& 
 10   &  Bonn    \\ \hline
\end{tabular} 

\medskip
Notes to Table 2.

$^1$ Seven VLBA antennas: SC, HN, NL, PT, OV, BR, MK

$^2$ Five LBA antennas: Parkes, Mopra, Compact Array (single antenna), 
Tidbinbilla, Hobart        

$^3$ Five LBA antennas: Parkes, Mopra, Compact Array (single antenna), 
DSS45, Hobart

$^4$ Five EVN antennas: Cambridge, Jodrell Bank, Noto, Medicina, Onsala

$^5$ Five EVN antennas: Effesberg, Jodrell Bank, Noto, Medicina, Onsala

\end{table*}

%

\begin{table*}
\caption{Details on the images}

\begin{tabular}{lccccccc} \hline
Source & $\nu$ & Restoring FWHM  & rms &
S$_{core} (VLBI) $ &S$_{tot} (VLBI) $ & S$_{PTI}$ (2.3GHz) & Figure \\
       &  GHz  & mas   & mJy b$^{-1}$ &
mJy   &   mJy    &   mJy & \# \\ \hline
3C17   & 4.99  &  $3.6\times1.4$ & 0.38 & 316 & 465  & 411 & 1 \\
PKS 0620$-$52& 2.29 & $33.5\times11.4$ & 0.95 &  80 &  80  & 123 & 2  \\
PKS 0625$-$35& 2.29 & $33.1\times13.7$ & 0.37 & 507 & 554  & 536 & 3  \\
PKS 1318$-$43& 2.29 & $30.0\times15.0$ & 0.86 & 353 & 407 & 411  & 4 \\ 
PKS 1333$-$33& 2.29 & $72.4\times19.9$ & 0.67 & 128 & 128  & 188 & 5  \\
3C317 & 1.66 & 233.0$\times$97.4 & 1.31 & 341 & 349  &  310 & 6  \\
3C317 & 4.99 & 57.9$\times$43.9  & 0.28 & 355 & 359  &   -  & 7   \\ 
3C317 & 4.99 & $5.0\times2.0$ & 0.33    & 296 & 360  &    - & 9  \\ \hline
\end{tabular} 
\end{table*}

\section{Comments on individual sources}

We report the images of the six radio galaxies 
in Figures \ref{fig:3c17} through \ref{fig:3c317_6h},
and describe them in detail in the following subsections.
In agreement with the  results found in the literature
on the parsec-scale morphologies of radio galaxies,
our objects are either dominated by a nuclear
component with asymmetric jet emission, or show only
marginal extension. In Table 3 we give the
observational parameters of the images, i.e. the total VLBI flux
density and the flux density in the most compact component,
assumed to be the core. For the unresolved sources these
two values are coincident.
For comparison we report also the 2.3 GHz core flux
density measured with the Parkes-Tidbinbilla real-time interferometer
(PTI, Morganti et al. 1997), while the flux density of the arcsecond scale
nucleus at 5 GHz (Morganti et al. 1993) is given in Table 1. 


\subsection{0035$-$02 (3C17)}

This is the most powerful and most distant object in our
sample. It is classified as a broad line radio galaxy (Tadhunter
et al. 1993) and it is also characterised by optical
polarisation (Tadhunter et al. 1997). High resolution 5 GHz 
VLA observations (Morganti et al. 1999) show that the source
is very asymmetric on the arcsecond scale both in
shape and brightness. 
The south-east arcsecond scale jet is aligned in position angle
(p.a.) $\sim +100^{\circ}$, and it is 
characterised by high brighness knots.
It sharply bends by $\sim +200^{\circ}$  at $\sim 10$ arcsec from 
the core. No jet is visible on the north-western side, but only a diffuse
lobe, with a ring-like structure, at $\sim$ 5 arcsec in projected 
distance form the core.
ROSAT X--ray emission was detected by Siebert et al. (1996).

\begin{figure}
\epsfysize=5cm
\epsfxsize=\hsize
\epsfbox{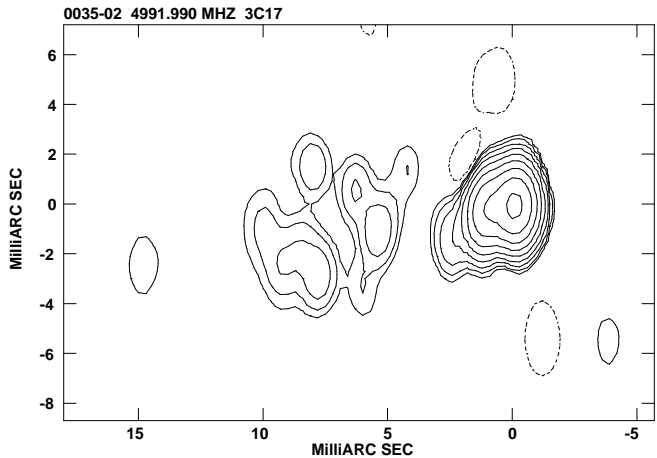}
\caption[]{5 GHz image of 3C17.
Peak is 275 mJy/beam. Contours are 0.9 mJy/beam $\times$
(-1, 1, 2, 4, 8, 16, 32, 64, 128, 256). 
Beam FWHM: 2.0 $\times$ 1.0 (mas) at 0$^{\circ}$. For this source 
1 mas $\sim$ 4 pc.}
\label{fig:3c17}
\end{figure}

Our parsec scale image, given in Figure \ref{fig:3c17},
shows a one-sided morphology 
oriented in east-west at $\sim +100^{\circ}$, 
roughly in agreement with the initial 
orientation of the south-east arcsecond scale jet.
Using the AIPS task JMFIT on this image, we performed a 
multi-component gaussian fit of the inner 5 mas.
The source can be described with an unresolved component, the 
westernmost one, which we identify with the core of the 
radio emission on the basis of its compactness,
and two more compact components. The component parameters 
derived from our fit are given in Table 4, where we report 
the flux S, size and position angle (p.a.) of each component,
the distance and position angle (P.A.) from the core for
the secondary components along the parsec-scale inner jet,
listed in order of increasing distance from the core.

\subsection{PKS 0620$-$52}

This radio source is associated with the brightest member of a galaxy cluster,
and its arcsecond scale morphology is typical of this large scale environment,
since it exhibits a wide-angle tail morphology with two asymmetric tails
(Morganti et al.  1993).  Very little optical information is available
for this source and its surroundings.  BeppoSAX X--ray observations were not
conclusive with respect to the existence of a non-thermal X--ray component of
nuclear origin (Trussoni et al.  1999). 

\begin{figure}
\epsfysize=5cm
\epsfxsize=\hsize
\epsfbox{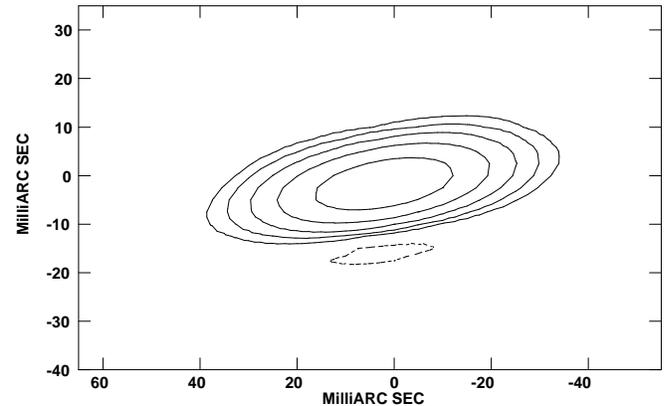}
\caption[]{2.29 GHz image of PKS 0620$-$52.
Peak is 80 mJy/beam. Contours: 3.0 mJy/beam $\times$
(-1, 1, 2, 4, 8, 16, 32, 64, 128, 256). 
Beam FWHM: 33.5 $\times$ 11.4 (mas) at $-84^{\circ}$.
For this source 1 mas $\sim$ 1.4 pc.}
\label{fig:0620}
\end{figure}

PKS 0620$-$52 is the weakest source in our VLBI sample and it is 
unresolved by the present 2.29 GHZ observations. The image is shown in
Figure \ref{fig:0620}. Assuming that flux density variations in this
source are negligible, comparison with the 
2.3 GHz PTI flux (see Table 3), suggests that extended flux, not
imaged by our observations, must be present on the sub-arcsecond scale.
The spectrum of the source in the range 2.3 - 5 GHz is typical
of gigahertz peaked sources (GPS).

\subsection{PKS 0625$-$35 (OH 342)}

The arcsecond scale emission for this radio galaxy is dominated
by a bright core, with a one-sided jet in p.a. $\sim +160^{\circ}$,
embedded in a low brightness halo (Simpson 1994), which
extends out to $\sim 4^{\prime}$  in the low resolution
image of Ekers et al. (1989).
X--ray BeppoSAX observations (Trussoni et al. 1999) are in agreement 
with the presence of  a non-thermal hard X--ray component, likely 
to be of nuclear origin. The X--ray emission from the central region
appears unabsorbed. 
These observational properties could be interpreted
either as the source being 
oriented at a small angle to the line of sight or that it is not
heavily affected by obscuration from a circumnuclear torus.

\begin{figure}
\epsfysize=5cm
\epsfxsize=\hsize
\epsfbox{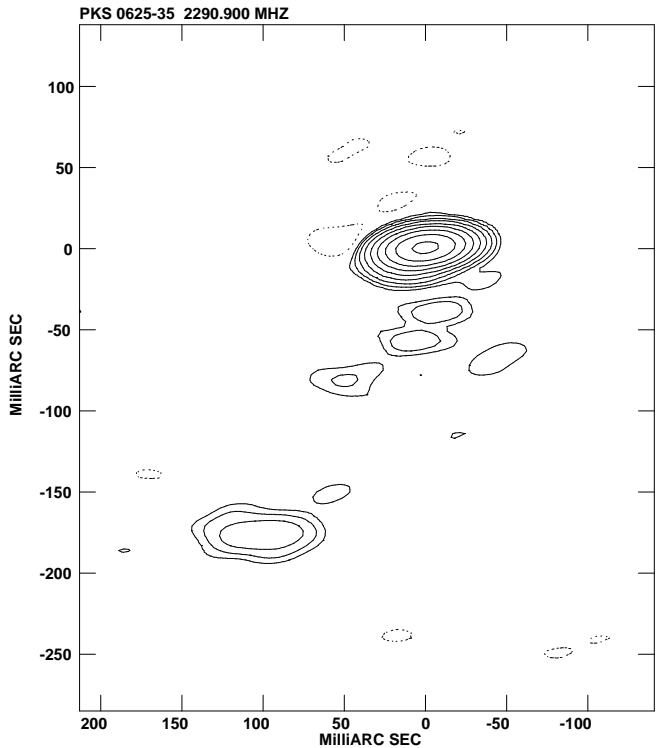}
\caption[]{2.29 GHz image of PKS 0625$-$35.
Peak is 490 mJy/beam. Contours: 1.2 mJy/beam $\times$
(-1, 1, 2, 4, 8, 16, 32, 64, 128, 256). 
Beam FWHM: 33.1 $\times$ 13.7 (mas) at $-84.3^{\circ}$.
For this source 1 mas $\sim$ 1.5 pc.}
\label{fig:0625}
\end{figure}

Our VLBI image, shown in Figure \ref{fig:0625}, is in agreement
with its arcsecond scale structure. It is dominated by a strong 
central component, which contains $\sim 90$\% of the total
VLBI flux density, with a faint jet aligned in p.a. 
$\sim 150^{\circ}$. The flux density given in Table 3 is
consistent with the PTI 2.3 GHz value if we account for the 
large uncertainties in the flux density scale for the LBA observations (see
Section 3). 

From Figure \ref{fig:0625} and from another image restored 
with a slightly super-resolved restoring beam in p.a. 0$^{\circ}$
(not shown here) it is clear that the dominant component is
resolved and shows
an extension in p.a. $\sim +130^{\circ}$. With the AIPS task
JMFIT we resolved it into two components. Details on the
fitting parameters are given in Table 4.
The southernmost jet component visible in Fig. \ref{fig:0625} is
very diffuse. It is located at $\sim +200$ mas, and has a total 
flux density of $\sim 28$ mJy. 


\begin{table*}
\caption{Results from the gaussian fit}

\begin{tabular}{lclrcccr} \hline
       &  &  &  &  &  &  &  \\
Source & $\nu$ & Component &  Flux & Size  & p.a. & 
Dist. from core & P.A. \\ 
       &  GHz &  & mJy & mas &  & mas  &  \\ \hline
3C17 & 4.99  &  Core & 316 & $<0.54$ & -  &  &  \\
     &       &  1st  & 113 & $<0.63$ & -  & 1.19 & 95$^{\circ}$  \\
     &       &  2nd  &   8 & $<0.74$ & -  & 2.96 & 110$^{\circ}$ \\
PKS 0625$-$35 & 2.29 & Core & 507 & 4.1$\times$ 1.0 & 161.3$^{\circ}$ & & \\
               &     & 1st  &  20 & $<8.3$ & - & 19.2 & 
130$^{\circ}$ \\
PKS 1318$-$43 & 2.29 & Core & 353 & $<4.6$  & - & & \\
              &      & 1st  &  53 & $<10.5$ & - & 18.8 & 
50$^{\circ}$ \\ \hline
\end{tabular} 
\end{table*}

\subsection{PKS 1318$-$43 (NGC 5090)}

This source is a typical FRI radio galaxy (Morganti et al. 1993; 
Lloyd et al. 1996), with symmetric arcsecond scale jets
and a total extension of $\sim$ 17 arcmin. The most prominent
feature of the arcsecond scale jets is their 
strong S-bend shape, from the region closest to the core  out 
to the end of the jets (Lloyd et al. 1996). One
possible explanation for this interesting structure comes
from the local environment of the optical counterpart,
which is believed to be interacting with the nearby 
spiral galaxy NGC 05091 (Smith \& Bicknell 1986). 
HI absorption has been detected
(using the VLA) against the core of this galaxy.  The HI profile appears
double peaked and the  
HI velocity derived is very  close to the systemic velocity of the galaxy 
(v = 3421 km s$^{-1}$).

\begin{figure}
\epsfysize=5cm
\epsfxsize=\hsize
\epsfbox{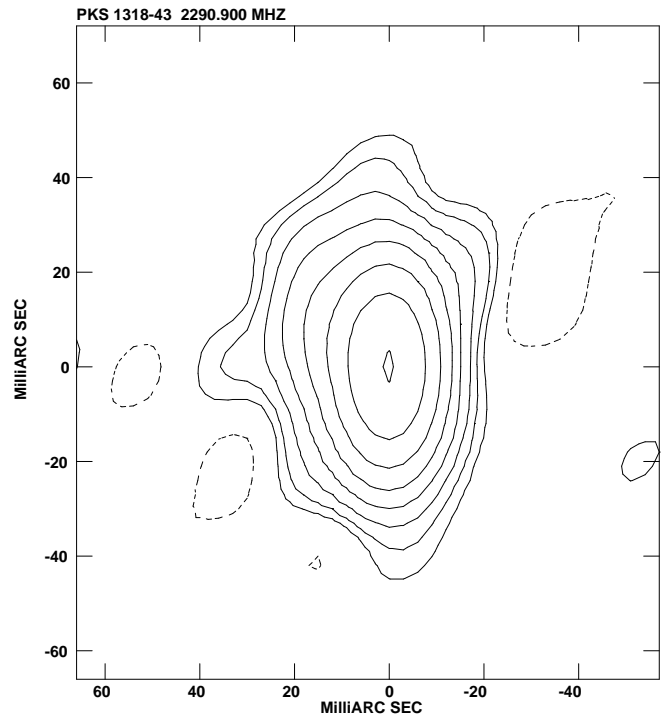}
\caption[]{2.29 GHz image of PKS 1318$-$43.
Peak is 345 mJy/beam. Contours: 2.6 mJy/beam $\times$
(-1, 1, 2, 4, 8, 16, 32, 64, 128, 256). 
Beam FWHM: 30$\times$15 (mas) at $0^{\circ}$. For this source 1 mas
$\sim$ 0.3 pc.}
\label{fig:1318}
\end{figure}

The parsec-scale morphology of PKS 1318$-$43 is shown in Figure 
\ref{fig:1318}. This image was obtained including only
the baselines in the range 0 - 8.5 M$\lambda$, given the
very noisy data in the longest baselines (due to the lower
sensitivity of the array) and it was convolved
with a restoring beam oriented in p.a. $0^{\circ}$, so as to better
show its extension. From Figure \ref{fig:1318} it is
clear that the source is resolved, and we 
fitted it with two componens, i.e. a dominant compact one,
which we assume to be the radio core, and a secondary peak 
at $\sim$ 18.8 mas, in p.a. $\sim +50^{\circ}$, aligned
with the northern arcsecond-scale jet. Details on the fit
are reported in Table 4.

The total flux density on the VLBI scale (see Table 3) is in
good agreement with the 2.3 GHz PTI data. Our value, 
coupled with the radio data
for the arcsecond radio nucleus of  PKS 1318$-$43 collected
from the literature (Table 3 and Lloyd et al. 1996), suggest
that the radio nucleus is self-absorbed at least up
to $\nu~=~8.6$ GHz.

\subsection{PKS 1333$-$33 (IC 4296)}

PKS 1333$-$33 is a very large radio galaxy (Killeen et al. 1986),
with radio power typical of transition objects between FRI and
FRII sources (see Table 1). Its radio morphology reflects the radio power,
and it is also intermediate between these two classes.
The source is characterised by an arcsecond self-absorbed nucleus,
which peaks at $\nu \sim 15$ GHz, and by two symmetric knotty jets 
which culminate in two ``warm'' spots at a distance of 
$\sim 12^{\prime}$ from the core, i.e. $\sim 250$ kpc.
The associated optical counterpart is a nearby elliptical galaxy,
with a featureless optical spectrum. X--ray observations
carried out with ASCA and BeppoSAX (Fiore et al. in preparation),
revealed the existence of a hard X--ray excess over the thermal
stellar component, most likely of nuclear origin, which led
them to propose that a low efficiency accretion flow
is hosted in the nuclear region of the galaxy.

\begin{figure}
\epsfysize=5cm
\epsfxsize=\hsize
\epsfbox{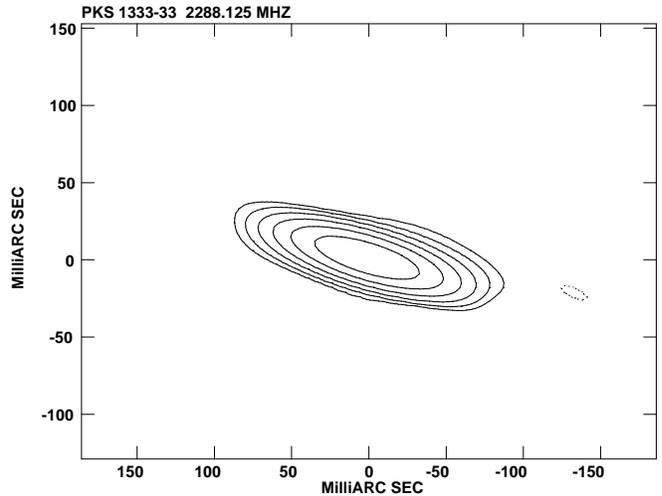}
\caption[]{2.29 GHz image of PKS 1333$-$33.
Peak is 119 mJy/beam. Contours: 2.0 mJy/beam $\times$
(-1, 1, 2, 4, 8, 16, 32, 64, 128, 256). 
Beam FWHM: 72.4 $\times$ 19.9 (mas) at $72.1^{\circ}$. For this source
1 mas $\sim$ 0.4 pc.}
\label{fig:1333}
\end{figure}

PKS1333$-$33 is almost unresolved at the resolution and
sensitivity of the present 2.29 GHz observations. 
As for PKS 1318$-$43,  due to sensitivity limits,
the very noisy data points on 
the longest baselines forced us to 
taper the u-v coverage in order to get a reliable image,
thus degrading the FWHM. Deconvolution of the
image presented in Figure \ref{fig:1333}, carried out with
the program JMFIT in the AIPS package, shows that it can be fitted
with a gaussian component of the size of $\sim$ FWHM/4.
The flux density value given in Table 3 is in good  agreement
with the arcsecond core spectrum given in Killeen et al. (1986)
and with the PTI 2.3 GHz data (Morganti et al. 1997),
allowing for the different resolution of our VLBI image. 
However, comparison with the PTI data confirms our estimate that 
the uncertainties on the absolute flux density scale may be of the order of 
$\sim$ 10 - 15\%.

\subsection{3C317}

3C317 is one of the best studied radio sources in our VLBI sample.
It is associated with the cD galaxy UGC 9799,
at the centre of the cooling flow cluster of galaxies A2052.
Its radio power is typical of FRI radio galaxies.
The source was imaged with the VLA in a wide range of
frequencies and studied in detail by Zhao et al. (1993).
According to its morphology it can be classified as 
core-halo, in that it is characterised by a compact region 
surrounded by a steep spectrum low brightness radio halo. 
It has an angular extent $\sim 1.5^{\prime}$ ($\sim$ 100 kpc)
and it is all embedded within the optical galaxy. 
The inner arcsecond scale region
of 3C317 is characterised by a counter-clockwise bent morphology, 
in a spiral-like 
shape. The source is also characterised by high integrated 
rotation measure, i.e. $-800$ rad m$^{-2}$, which can be 
explained assuming that a magnetised 
high density medium is present around  
the radio emission (Taylor, Inoue \& Tabara 1992). The polarisation
properties and rotation measure structure on the arcsecond scale
were studied in detail by Owen \& Ge (1994). Their images also
revealed that the bending region, well visible in Zhao et al.
(1993) is not a jet, in that it turns out to be a diffuse and ``blobby''
emission when observed at higher resolution.
In the optical band 3C317 shows several emission line filaments, cospatial
with the radio emission (Burns 1990) and its optical spectrum
is typical of FRI radio galaxies, i.e. low ionisation and dominant
stellar continuum (Tadhunter et al. 1993).

\begin{figure}
\epsfysize=5cm
\epsfxsize=\hsize
\epsfbox{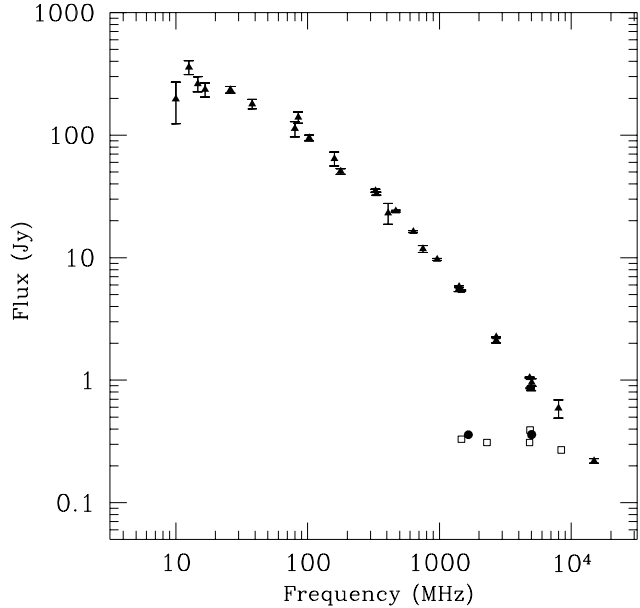}
\caption[]{Spectrum of 3C317 from 10 MHz to 15 GHz. Filled triangles
refer to the total flux density; filled circles the VLBI data
presented in this paper; open squares are arcsecond core flux density 
data.}
\label{fig:3c317_spix}
\end{figure}

In order to get a clearer picture on the nature of this source 
we collected all the total flux density data available in the literature 
and plotted them in figure Figure \ref{fig:3c317_spix}
(filled triangles), together with the parsec-scale data
we derived in this paper and the values for the arcsecond
scale core published in Zhao et al. (1993) and Morganti et al. (1993).
Most of the total flux density data were taken from the K\"uhr
Catalogue of Radio Sources (1979). The values at 10, 12.6, 14.7 and 16.7 
MHz were taken from Braude et al. (1979). 
From a few MHz up to $\sim 8.4$ GHz
the total flux density of the source is clearly dominated by
the presence of the steep spectrum radio halo ($\alpha_{0.1 GHz}^{8.4 GHz} 
\sim 1.6 \pm 0.1$, Zhao et al. 1993).

\begin{figure}
\epsfysize=5cm
\epsfxsize=\hsize
\epsfbox{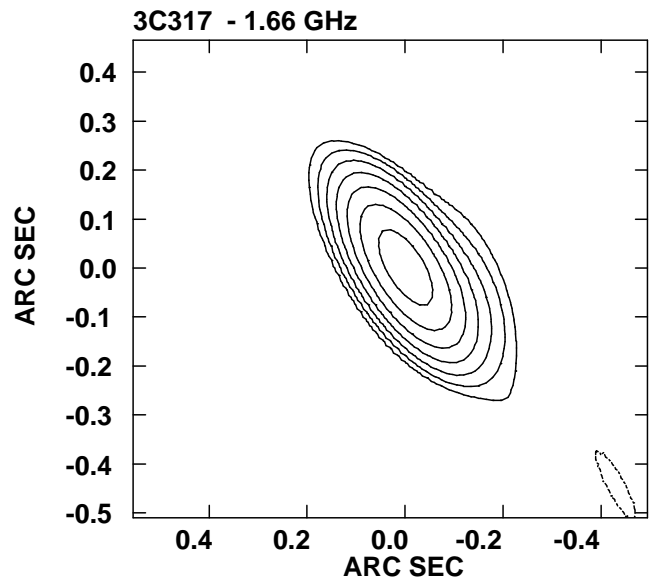}
\caption[]{1.66 GHz MERLIN-only image 0f 3C317.
Peak: 341 mJy/beam. Contours: 3.6 mJy/beam $\times$
(-1, 1, 2, 4, 8, 16, 32, 64, 128, 256). 
Beam FWHM: 233.0 $\times$ 97.4 (mas) at $31.2^{\circ}$. Linear scale:
1 mas $\sim $ 1 pc.}
\label{fig:3c317_18mer}
\end{figure}

\begin{figure}
\epsfysize=5cm
\epsfxsize=\hsize
\epsfbox{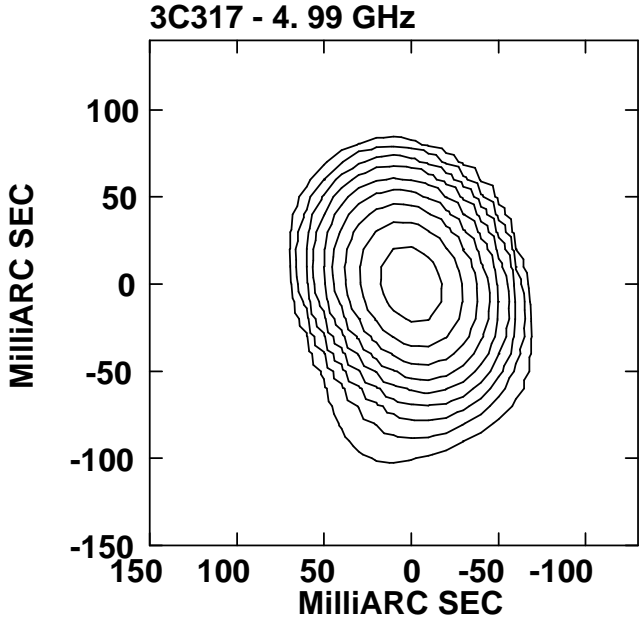}
\caption[]{4.99 GHz MERLIN-only image 0f 3C317.
Peak: 353 mJy/beam. Contours: 0.9 mJy/beam $\times$
(-1, 1, 2, 4, 8, 16, 32, 64, 128, 256). 
Beam FWHM: 57.9 $\times$ 43.9 (mas) at $24.2^{\circ}$. Linear scale:
1 mas $\sim $ 1 pc.}
\label{fig:3c317_6mer}
\end{figure}

\begin{figure}
\epsfysize=5cm
\epsfxsize=\hsize
\epsfbox{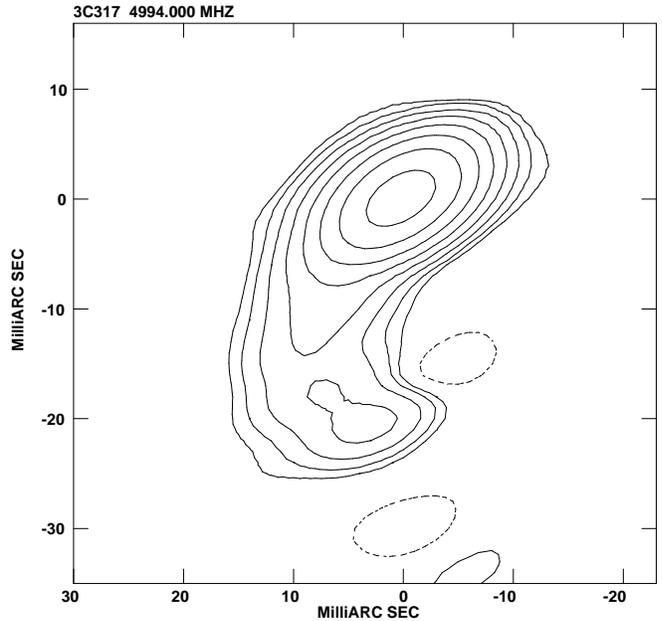}
\caption[]{Full resolution 4.99 GHz EVN+MERLIN image of 3C317.
Peak is 216 mJy/beam. Contours: 1.2 mJy/beam $\times$
(-1, 1, 2, 4, 8, 16, 32, 64, 128, 256). 
Beam FWHM: 8.9 $\times$ 4.2 (mas) at $-70.8^{\circ}$. For this source
1 mas $\sim$ 1 pc.}
\label{fig:3c317_6l}
\end{figure}

\begin{figure}
\epsfysize=5cm
\epsfxsize=\hsize
\epsfbox{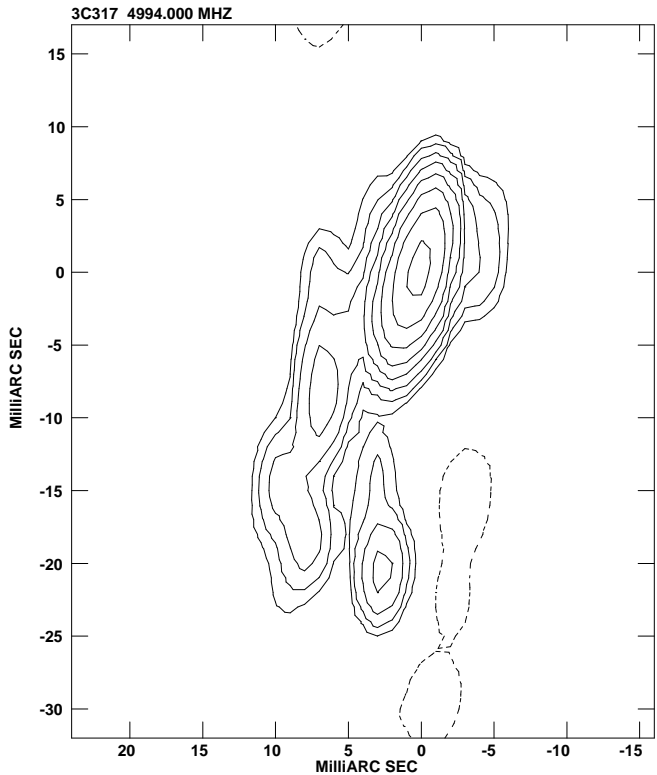}
\caption[]{4.99 GHz EVN+MERLIN image of 3C317 convolved with
FWHM: 5 $\times$ 2 (mas) at $0{\circ}$.
Image peak is 165 mJy/beam. Contours are 1.0 mJy/beam $\times$
(-1, 1, 2, 4, 8, 16, 32, 64, 128, 256).}
\label{fig:3c317_6h}
\end{figure}

Given the arcsecond morphology of 3C317, we chose to observe 
this source with the EVN+MERLIN at 1.66 GHz and 4.99 GHz,
so as to get good sensitivity and u-v coverage at 
short spacings and ensure  
the detection of low brightess extended emission on the 
sub-arcsecond scale, if present. The images we obtained are shown in
Figures \ref {fig:3c317_18mer} to \ref{fig:3c317_6h} and the
details on the images are given in Table 3. Figure 
\ref{fig:3c317_6h} was obtained with the same u-v coverage as Figure 
\ref{fig:3c317_6l}, but higher weight was given to the long baselines,
in order to better show
the structure of the dominant component in Figure \ref{fig:3c317_6l}. 


MERLIN-only images of 3C317 at 1.66 GHZ and 4.99 GHz are 
shown in Figures \ref{fig:3c317_18mer} and \ref{fig:3c317_6mer}.
The source is barely resolved at both frequencies and resolutions,
with an extension in p.a. $\sim -60^{\circ}$ at 1.66 GHz, and 
p.a. $\sim 170^{\circ}$ at 4.99 GHz.
The extension in different position angles at different 
frequencies and resolutions suggests that the source may
have a complex morphology.
This is confirmed by our 4.99 GHz full resolution
EVN+MERLIN image.
At milliarcsecond resolution the source is dominated by a 
compact component and a 
jet, bending at $\sim 20$ mas from the nucleus (i.e. $\sim 20$ pc).
The images shown in Figures \ref{fig:3c317_6l} and \ref{fig:3c317_6h}
are very similar to the arcsecond scale ones presented in Zhao et al.
(1993), i.e. the jet bends in an anti-clockwise direction on
both scales. We present no EVN+MERLIN image at 1.66 GHz since the
very noisy EVN data in this observation do not add any information to
the MERLIN-only image.

The data presented in this paper are plotted in Figure 
\ref{fig:3c317_spix} together with the PTI data (see Table 3) and
the high resolution 8.4 GHz data published in Zhao et al. (1993).
Our VLBI data give a spectral index 
$\alpha_{1.66}^{4.99} \sim 0$ for the total VLBI structure, 
in agreement with the arcsecond core spectrum
derived by Zhao et al. (1993). 

\noindent
On the basis of our observations and from inspection of the plot 
shown in Figure
\ref{fig:3c317_spix} we estimate that the core turnover frequency
is $\sim$ 5 GHz, however the scatter in the VLA literature data
(Morganti et al. 1993; Zhao et al. 1993)
for the core does not rule out a flat spectrum, i.e. $\alpha \sim 0$,
from 1.6 GHz to 8.4 GHz.
We conclude therefore that the steep spectrum radio halo 
``hides'' an active nucleus with flat spectrum.

From Figure \ref{fig:3c317_6h} a short counter-jet feature is 
visible on the opposite side of the main parsec-scale jet, which
persisted even though we did not include it in the self-calibration
process. 
We cannot rule out the possibility that it is due to a
residual calibration error, however this morphology could also
be interpreted in terms of Doppler boosting, as for the other
sources presented in this paper (see next Section).

\section{Discussion and conclusions}
 
Two out of the six radio galaxies from the 2-Jy sample
presented in this paper, i.e. PKS 0620$-$52 and PKS 1333$-$33,
are unresolved at the sensitivity and resolution of the present 
observations, despite their extended large scale morphology.
The extension revealed by a gaussian 
fit on PKS 1333$-$33 suggests that some structure exists
on the parsec scale, which higher sensitivity observations
will probably allow to image.

\medskip
The remaining four sources show an asymmetric morphology,
which could be explained assuming Doppler boosting 
effects in an intrinsically symmetric source, as from radio
loud unified models.

In order to derive estimates for the intrinsic plasma
speed $\beta = {v \over c}$ and limits to the viewing angle 
of the radio source with respect to the line of sight $\theta$,
we used two quantities derived from the observations:

\noindent 
{\it (a)} the brightness ratio between the parsec-scale 
jet and counterjet defined as R=${S_J} \over {S_{CJ}}$;

\noindent 
{\it (b)} the 5 GHz arcsecond scale core dominance with respect
to the source total power at 408 MHz (see Table 1), measured comparing
the 5 GHz core power logP$_c$ for the sources in our sample to the
well known observational relation between logP$_c$ (5 GHz) and 
logP$_{tot}$ (408 MHz) in low power radio galaxies scaled for
our choice of H$_o$ (see Giovannini et al. 1994 for details). 
In the light of Doppler boosting, the difference between
the observed 5 GHz core power and the core power
derived from the correlation, assumed to be the core power
for a source oriented at 60$^{\circ}$ from the line of sight,
is entirely due to the different orientation of the radio
plasma to the line of sight. As in Giovannini et al. (1994)
we define K=$\sqrt{{P_c(\theta)}\over{P_c(60^{\circ})}}$.

In Table 5 we report the limits to the boosting parameters 
we derived applying these two methods. The ratio R was 
computed at the distance of the first component reported in
Table 4. 
Since no total flux density measurement is available at 408 MHz
for PKS 1318$-$43, we assumed a value of 16 Jy, derived from
extrapolation of the Molonglo 843 MHz flux density with a
spectral index $\alpha$ = 1. 

\begin{table}
\caption{Doppler boosting parameters}

\begin{tabular}{lrrrccc} \hline
       &                          &                &               \\
Source & R & $\theta_{max}$ & $\beta_{min}$ & K & $\theta_{max}$ &
$\beta_{min}$ \\
       &                          &               &               \\ \hline
3C17   & $\ge~50$  &  $49^{\circ}$ & $0.65$ & 
6.31   & $ 33^{\circ}$ & $0.84$ \\
PKS 0620$-$52 &   &    &   & 2.51 &  $53^{\circ}$ & $0.60$ \\
PKS 0625$-$35 & $\ge~12.5$  &  $61^{\circ}$ & $0.47$ &
3.80   & $43^{\circ}$ & $0.74$ \\ 
PKS 1318$-$43   & $\ge~8$  &  $67^{\circ}$ & $0.40$ &
2.54   & $53^{\circ}$ & $0.60$   \\
3C317   & $\sim 4$  &  $68^{\circ}$ & $0.36$ &
6.68 & $32^{\circ}$ & 0.85 \\ \hline
\end{tabular} 
\end{table}

Even though the core dominance leads to more stringent
limits for $\beta$ and $\theta$, the values we derived
with these two methods are in reasonable agreement with
each other, except for 3C317, which we will
discuss in further detail below.

\medskip
3C17 shows the highest degree of asymmetry among our sources,
consistent with the
fact that it is a broad line radio galaxy, hence expected 
to be oriented at small to moderate angles to the line of sight.

For PKS 0620$-$52 the core dominance suggests that the plasma
speed in this source can be mildly relativistic to relativistic.

PKS 0625$-$35 and PKS 1318$-$43 are both 
one-sided on the parsec-scale, but their degree of
asymmetry is less severe than for 3C17 and leads to 
upper limits to the viewing angle fully consistent
with the expectations from unification.
In particular, for PKS 0625$-$35 the upper limit
derived for $\theta_{max}$
with both methods is in very good  agreement with the independent 
estimate done by Trussoni et al. (1999).
For this source, the upper limit found, coupled with the indication 
of a hard X--ray unobscured nuclear component, argues in favour 
of the fact that nuclear obscuration from a torus is probably 
not relevant in this object.

Interestingly enough, for PKS1333$-$33 the core power at 5 GHz 
derived on the basis of the logP$_c$ - logP$_{tot}$ correlation is 
in perfect agreement with the measured value (see Table 1),
thus suggesting that the source is viewed at an angle 
$\theta \sim 60^{\circ}$.

\medskip
The case of 3C317 is more complex. 
Even though we cannot rule out the possibility that 3C317 is an 
edge-on radio galaxy with intrinsic bends on the parsec scale
amplified by a small viewing angle (see Table 5), an
alternative explanation for the properties of 3C317 should
also be taken into account. We suggest that the observed bent morphology 
is the result of the interaction  between the relativistic plasma 
and a very dense external medium. Under this second hypothesis, 
the parsec-scale jets are  disrupted in the sub-arcsecond region, 
as consequence of gas accretion in the optical counterpart UGC 9799
from the cluster cooling flow, thus preventing the source from
evolving into a classical FRI radio galaxy. 
The MERLIN-only images in Figure \ref{fig:3c317_18mer} and
\ref{fig:3c317_6mer}
at 1.66 GHz and at 4.99 GHz respectively show no hint
of jets or extended emission at the resolution of the order of
50 mas ($\sim$ 50 pc) at 5 GHz or 150 mas at 1.7 GHz, 
and this suggests that jet disruption
may take place at most at such distance from the core.  
Zhao et al. (1993) gave estimated the sonic radius $r_s$ for
3C317, i.e. the region where radio jets may become unstable
and disrupt as a consequence of the cooling flow at the centre
of the galaxy, and gave a value $r_s \le $0.4 kpc.
Our parsec-scale images of 3C317 first of all confirm that
the radio plasma is indeed collimated at least out to
a distance of $\sim$ 20 pc from the core, moreover they  
place a new upper limit to the sonic radius $r_s \sim$ 50 pc.

As shown in the previous Section, comparison between the
nuclear spectrum and the large scale spectrum of 3C317 
suggests that a very active nucleus is embedded in a steep
spectrum radio halo. If we make the simplified assumption
that the parsec-scale emission in the source is in
equipartition and if we make the standard equipartition
assumptions, i.e. cylindrical geometry, 
filling factor $\phi$ = 1, ratio of protons to electrons 
energy k=1, and integrate in the range of frequencies 
$10^7 - 10^{11}$ Hz, we obtain an estimate for the average
parsec-scale 
magnetic field of the VLBI image B$_{eq} \sim 3.5~\times~10^{-3}$ G.
From the core spectrum reported in Figure \ref{fig:3c317_spix} we 
estimated a break frequency $\nu_{\star} \ge 15$ GHz, and with
the derived value for B$_{eq}$ we obtain a radiative age 
$t_{syn} \le$ 2000 yrs. These numbers confirm 
that the nuclear region in 3C317 is active, furthermore
they suggest that the radiating electrons are very young.

\medskip
To summarise, the six radio galaxies presented in this
paper exhibit parsec-scale properties in agreement 
with the unified models for radio loud galaxies, i.e.
they are characterised by at least mildly relativistic
speeds and are viewed at intermediate to large angles to
the line of sight. 

For 3C17 and PKS 0625$-$35 our 
results are in very good agreement with independent estimates
derived on the basis of their optical and X-ray properties
respectively. 

Our detailed study on 3C317 shows that the source is
characterised by parsec-scale collimated jets, which disrupt
on the sub-kiloparsec region, possibly as consequence of
the cooling flow at the centre of the galaxy. Our spectral
analysis of the source core region suggests that the source
is very young.

\begin{acknowledgements}
We thank dr. Daniele Dallacasa for many insightful
discussions and critical reading of the manuscript.
Thanks are due to M. Tugnoli for his invaluable help
in running the Mk2 correlator in Bologna.
The Australia Telescope Compact Array (/Parkes telescope/Mopra
telescope/Long Baseline array) is part of the Australia Telescope
which is funded by the Commonwealth of Australia for operation
as a National Facility managed by CSIRO.

\end{acknowledgements}


\end{document}